\documentclass[pdflatex,sn-nature]{sn-jnl}% Default
\usepackage{graphicx}%
\usepackage{multirow}%
\usepackage{amsmath,amssymb,amsfonts}%
\usepackage{amsthm}%
\usepackage{mathrsfs}%
\usepackage[title]{appendix}%
\usepackage{xcolor}%
\usepackage{textcomp}%
\usepackage{manyfoot}%
\usepackage{booktabs}%
\usepackage{algorithm}%
\usepackage{algorithmicx}%
\usepackage{algpseudocode}%
\usepackage{listings}%
\usepackage{siunitx}
\theoremstyle{thmstyleone}%
\theoremstyle{thmstyletwo}%

\theoremstyle{thmstylethree}%

\begin{document}

%\title[Article Title]{Article Title}
%\title{The phase diagram of epithelial cells and tissues}
\title{Epithelial Tissues from the Bottom-Up: Contact Inhibition, Wound Healing, and Force Networks}

\author[1]{\fnm{Anshuman} \sur{Pasupalak}}
\author[1]{\sur{Zeng} \fnm{Wu}}
\author[1,2]{\fnm{Massimo} \sur{Pica Ciamarra}}\email{massimo@ntu.edu.sg}

\affil[1]{\orgdiv{Division of Physics and Applied Physics, School of Physical and Mathematical Sciences}, \orgname{Nanyang Technological University}, \orgaddress{\street{21 Nanyang Link}, \postcode{637371}, \country{Singapore}}}
\affil[2]{\orgdiv{Consiglio Nazionale delle Ricerche}, \orgname{CNR-SPIN}, \orgaddress{\city{Napoli}, \postcode{I-80126}, \country{Italy}}}

%%==================================%%
%% sample for unstructured abstract %%
%%==================================%%

\abstract{
In processes such as embryo shaping, wound healing, and malignant cell invasion, epithelial cells transition between dispersed phases, where the cells move independently, and condensed phases, where they aggregate and deform to close gaps, forming confluent tissues. Understanding how cells regulate these transitions and how these transitions differ from those of inert particles remains an open challenge. Addressing these questions requires linking the macroscopic properties of tissues to the mechanical characteristics and active responses of individual cells, driven by sub-cellular processes. Here, we introduce a computational model that incorporates key factors such as cell deformability, lamellipodium-driven dynamics, cell-junction-mediated adhesion, and contact inhibition of locomotion (CIL)—a process where cells alter their motion upon contact with others. We demonstrate how these factors, along with cell density, regulate the dynamical and mechanical properties of tissues. We show that CIL imparts unique living-like behaviors to cells and tissues by reducing density fluctuations. This reduction in fluctuations affects the dynamics: it inhibits cell motion in steady states but promotes it in the presence of gaps, accelerating wound healing. Furthermore, the stabilization of tensile states by CIL, which would otherwise fracture, enables the formation of tensile force chains.
}

%%================================%%
%% Sample for structured abstract %%
%%================================%%

%\keywords{keyword1, Keyword2, Keyword3, Keyword4}
%%\pacs[JEL Classification]{D8, H51}
%%\pacs[MSC Classification]{35A01, 65L10, 65L12, 65L20, 65L70}

\maketitle
Cell and tissue motion regulation is crucial for developing and sustaining life. 
During embryonic development, epithelial tissues protect the embryo and actively shape it through coordinated movements. 
In adulthood, they protect organs like the lungs, heart, and glands and set in motion during critical biological processes such as wound healing. 
Dysfunctions in the regulatory mechanisms governing cell and tissue locomotion are responsible for the invasiveness of malignant cells~\cite{Bissell2001,Ilina2020}.

Cells' traction forces, cellular density, adhesion, and active responses, such as contact inhibition of locomotion, regulate the motile properties of cell monolayers~\cite{Alert2020}. However, it remains unclear whether these factors, aside from traction forces~\cite{Garcia2015}, promote or inhibit motion. For instance, cellular crowding promotes the jammed phase by increasing kinetic constraints, akin to the behavior of inert colloidal particles, and high density is crucial for achieving a confluent jammed state. Yet, the role of crowding is experimentally debated~\cite{Park2015,Mitchel2020}, and the popular Vertex model, which describes a tissue as an evolving two-dimensional space tessellation, predicts that crowding promotes fluidity~\cite{su2016overcrowding}. Cell-junction-mediated adhesion forces provide tissues with tensile strength and promote the jammed phase~\cite{Garcia2015}. Indeed, experiments on cancer cells have shown that suppressing adhesion promotes the reverse unjamming transition~\cite{Bissell2001,Ilina2020}. However, some experiments~\cite{Park2015} and the Vertex model suggest that adhesion in confluent tissues promotes the unjammed phase by influencing the typical particle shape~\cite{Bi2015,malinverno_endocytic_2017,Lawson-Keister2021}.

Contact inhibition of locomotion (\textsc{cil}) is an active cellular response crucially influencing cellular motion~\cite{Abercrombie1970, Abercrombie1979, Carmona-Fontaine2008}.
Crawling cells exhibit front-rear polarity and move by extending lamellipodia protrusions at their leading edge, which adhere to surfaces and exert traction for movement~\cite{Alberts2017}.
\textsc{cil} refers to cells' ability to respond to contacts established by the front-located lamellipodia by altering their front-rear polarity, and hence their self-propelling direction. 
Indeed, \textsc{cil} is considered crucial for directing cellular motion in wound healing~\cite{Roycroft2016, Mayor2010, Carmona-Fontaine2008} and for the motion of cellular clusters~\cite{Copenhagen2018}.
While potential regulators of this biochemical response are known~\cite{Carmona-Fontaine2008, Roycroft2016, Scarpa2016},
how \textsc{cil} influences tissue macroscopic behavior is unclear.
On the one side, it should suppress molecular motion, as the name implies. 
Indeed, the dysfunction of \textsc{cil} has been linked to the invasiveness of malignant cells~\cite{Abercrombie1970, Abercrombie1979, Friedl2009, Friedl2018}.
On the other side, \textsc{cil} encourages tissue spreading and, when modelled as an alignment interaction between close cells~\cite{Zimmermann2016, Smeets2016a}, it induces collective superdiffusive motion at high density rather than inhibiting motion.
Numerical studies have shown that different plausible approaches to modeling \textsc{cil} can result in macroscopic tissues with qualitatively distinct properties~\cite{Camley2014,George2017}. Therefore, a detailed microscopic description of \textsc{cil} is crucial for understanding and rationalizing its influence on macroscopic tissue behavior~\cite{Camley2017}.

In this study, we investigate how crowding, self-propulsion, adhesion, and \textsc{cil} collectively influence the macroscopic properties of epithelial tissues using a novel high-resolution model for isolated cells and tissues. 
Our model's high spatial resolution is essential for accurately capturing particle deformability, interactions, lamellipodia-driven motility, and the effects of \textsc{cil}, providing access to both internal cellular forces and junction-mediated intercellular ones. 
We find that motile forces also induce a density-motility phase diagram that resembles that of colloidal suspensions with short-range attraction, featuring a gel phase at low densities and a glassy phase at high densities. 
In contrast, \textsc{cil} induces a nonequilibrium phenomenology that is qualitatively distinct from thermal systems. 
It dampens density fluctuations, reduces the extent of the coexistence region, promotes wound healing.
Furthermore, \textsc{cil} stabilizes tensile states that would otherwise cavitate by promoting the emergence of tensile force chains.
%Our results suggest interpreting wound healing as the \textsc{cil}-driven dynamical process through which tissues counteract density fluctuations, providing new insights into its dynamic and mechanical characteristics. 

\section*{NexTissUe: tissues with sub-cellular resolution}
\begin{figure}[t!]
\centering
\includegraphics[width=0.5\textwidth]{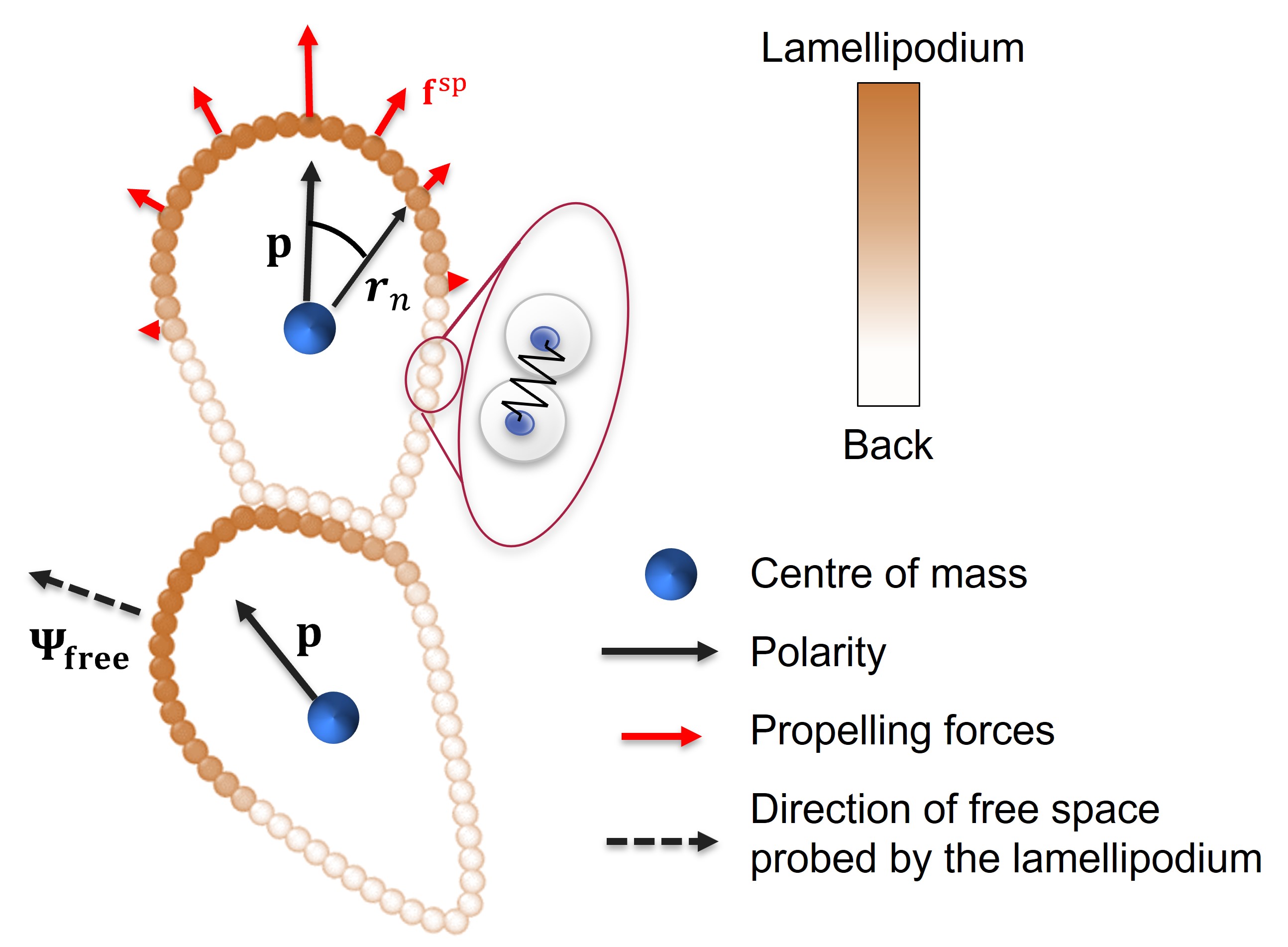}
\caption{
NexTissUe models a cell as a ring polymer with an area constraint and assigns polarity to its centre of mass. 
The polarity, along with the distance of each monomer from the centre of mass, defines the monomer's contribution to the lamellipodium, shown by the colour gradient. 
Monomers in the lamellipodium experience propelling forces that move the cell forward, as illustrated in red for the top cell. 
Contact inhibition of locomotion is modelled as the polarity’s tendency to orient towards free space probed by the lamellipodium, where the direction of free space is determined by the positions of the lamellipodium’s monomers that experience no interaction forces. 
This direction is illustrated for the bottom cell.
\label{fig:model}
} 
\end{figure}

We introduce \textsc{NexTissUe}, a model for adhesive, active and deformable cells~\cite{Anshuman_phdthesis} building on the two-dimensional (2D) description of deformable inert particles introduced in Ref.~\cite{Boromand2018}. 
Each cell is a ring polymer with an elastic energy constraint on the enclosed area, as illustrated in Fig.~\ref{fig:model}.
A cell's elastic energy is $u = \sum_i k_P(l_i-l_0)^2 + k_A(A-A_0)^2$.
The first term models the contractility of the actomyosin cortex, and it is a sum over the $N_m$ harmonic bonds of the ring polymer, $l_i$ being the length of the $i$-th bond and $l_0$ its rest length.
The second term penalises deviations of the area $A$ of the cell from the value $A_0$.
The elastic energy $u$ recalls that of the Vertex model and can be similarly non-dimensionalised via the introduction of a shape-parameter $p_0 = N l_0/\sqrt{A_0}$~\cite{Bi2015}.
However, contrary to the Vertex model, here $p_0$ is a single-cell property that does not affect cell-cell interactions. 

Cells tightly bind via the formation of transmembrane junctions.
The typical intermembrane space~\cite{Farquhar1963} at these junctions ($\approx 10$nm) is much smaller than a cell's typical linear size ($\approx 10\mu$m).
We model these features by assuming monomers of different cells interact via a potential comprising repulsion at short distances and an attractive well extending a few per cent of a cell's linear size, $d_0 \approx 2\sqrt{A_0/\pi}$ as detailed in the Supplementary Information (SI) Fig.~S1.
We use many monomers to describe each cell, ensuring the interaction force between adjacent cells is normal to their surface of separation~\cite{Boromand2018}.
Henceforth, we are effectively modelling sticky, frictionless, deformable colloidal particles.

The front-located lamellipodium drives cell motility and \textsc{cil}.
We define the lamellipodium by associating a polarity versor ${\bf p}= (\cos\theta,\sin\theta)$ with the cell's centre of mass, and characterizing each monomer with a scalar $\Lambda_n = \max(0,{\bf p}\cdot \hat {\bf r}_n) \in [0,1]$, ${\bf r}_n$ being its distance from the centre of mass. 
For monomers in the leading edge, $\Lambda_n > 0$, while for the others, $\Lambda_n = 0$. 
A self-propelling force ${\bf f}_n^{\rm sp} = f^{\rm sp} \Lambda_n \hat {\bf r}_n$ acting on each monomer reproduces the cell's lamellipodium-driven motility, as illustrated in Fig.~\ref{fig:model}.
We incorporate dissipation by adding a viscous drag force proportional to their velocity $-\gamma {\bf v}_n$, working in the overdamped limit.

Cell polarity evolves through stochastic fluctuations and \textsc{cil}.
By inhibiting the extensions of lamellipodia towards existing contacts~\cite{Abercrombie1954,Carmona-Fontaine2008, Mayor2010, Copenhagen2018, Zimmermann2016, Camley2016},  \textsc{cil} favours motion towards the free space {\it probed by the lamellipodium}.
We thus model \textsc{cil} as an alignment interaction of a cell polarity towards ${\bf \Psi}_{\rm free}=(\cos \Psi_{\rm free},\sin \Psi_{\rm free})$, which we identify by exploiting the high spatial resolution of our model as described in the SI Fig.~S2.
\textsc{cil} plays no role for isolated cells, which have no contacts, and for cells surrounded by other cells, which have no adjacent free space.
We model this \textsc{cil}'s dependence on the cell's vicinity by assuming it has a strength $\beta=f_c(1-f_c)$, with $f_c$ the fraction of monomers of a cell interacting with other cells.
Overall, cell polarities evolve according to
\begin{equation}
\frac{d\theta}{dt} = \sqrt{2D_r} \xi - \frac{\beta}{\tau_{\textsc{cil}}} \sin(\theta-\psi_{\rm free}),
\label{eq:polarity}
\end{equation}
where $D_r$ is a rotational diffusion coefficient, and $\xi$ is Gaussian white noise of unit variance.
We highlight \textsc{cil} is a non-reciprocal cell-cell interaction.

We perform simulations of $N=1000$ cells, each comprising $50$ monomers, using periodic boundary conditions. 
We use a cell linear size, $d_0 \approx 20 \mu m$, as a unit of length, and measure the self-propelling forces in terms of the minimal one needed to detach two adherent cells, $f_{\rm ad} \simeq 10^{-7}\;\unit{\newton}$ ~\cite{Esfahani2021} (SI Fig.~S3). 
The corresponding velocity scale is $v_0=f_{\rm ad}/\gamma=10\;\unit{\um.h^{-1}}$.
Cells have a persistence length $v_0D_r^{-1}$ comparable to their linear size~\cite{malinverno_endocytic_2017}, and hence a Peclet number of the order of one. 
This fixes our time unit, $D_r^{-1} \simeq 30 \min$.
Cell repolarization involves remodelling protein assemblies within the cell, a process occurring~\cite{Davis2015,Wyatt2016} on a much shorter time scale $\tau_{\textsc{cil}}$.
Henceforth, we expect $\tau_{\textsc{cil}} D_r \ll 1$. 
We also consider the opposite limit $\tau_{\textsc{cil}} D_r \gg 1$, where $\textsc{cil}$ plays no role, to investigate how contact inhibition influences tissues' macroscopic properties.
The parameter $p_0$ entering the energy function of an isolated cell characterizes its deformability. 
We fix $p_0=3.9$, a value at which non-adhesive cells jam at confluency on increasing the number density~\cite{Boromand2018}.
We measure crowding via the effective volume fraction $\phi_{\rm eff}=\rho A$, with $\rho$ the number density and A the area of non-motile cells in dilute conditions. 
If cells do not shrink or swell, then $\phi_{\rm eff} \simeq 1$ at confluency.

\section*{Motility-density phase diagram}
\begin{figure*}[t!]
\centering
\includegraphics[width=1\textwidth]{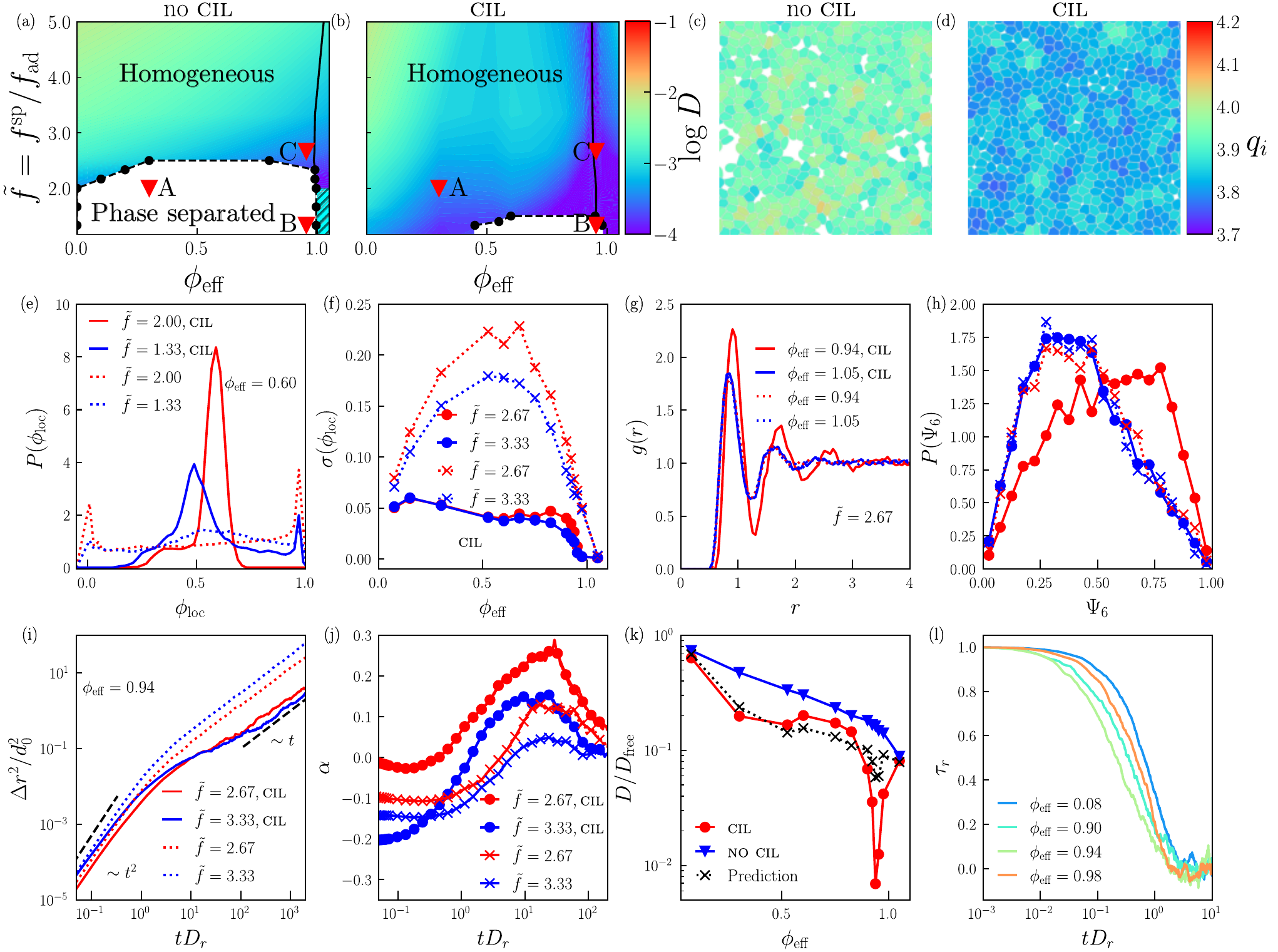}
\caption{
Panels (a) and (b) illustrate cell phase diagrams in the motile force strength $\tilde f$ and effective volume fraction $\phi_{\rm eff}$ plane, in the absence (a) and with (b) \textsc{cil}. The white area denotes the coexistence region where the system phase separates into a low-density and a high-density phase.
The colour code reflects the particles' diffusion coefficient. 
In the striped region at low $\tilde f$ and high $\phi_{\rm eff}$ in (a), the dynamics is too slow for an accurate measure of the diffusion coefficient.
Panels (c) and (d) illustrate a fraction of the investigated system at the state point indicated with C in (a) and (b), with particles coloured according to their shape index $q_i=P_i/\sqrt{A_i}$, with $A_i$ and $P_i$ particle and perimeter. 
The colour difference demonstrates that \textsc{cil} swells the particles, reducing $q_i$ and promoting confluence, as visually apparent.
Panel (e) illustrates the distribution of the local volume fraction that becomes bimodal at small activities. 
The peak positions identifying the coexisting volume fractions correspond to the full circles in (a) and (b).
Panel (f) illustrates the density dependence of the local volume fraction's standard deviation and demonstrates that \textsc{cil} suppresses density fluctuations.
Panel (g) shows that \textsc{cil} does not influence the radial distribution function at high density. In contrast, it shifts the position of the first peak at lower ones, acting as an effective repulsion. This effective repulsion also induces changes in the distribution of the hexatic order parameter (h).  
(i) and (j) illustrate the time dependence of mean square displacement and non-Gaussian parameter.
Panel (k) illustrates the effective volume fraction dependence of the diffusivity. It further shows that \textsc{cil}'s diffusivity can be predicted by assuming that $\textsc{cil}$ primarily affects the dynamics by reducing the polarity correlation time, which we extract from the polarity-polarity correlation functions illustrated in (l).
Except for the local volume fraction, we evaluate structural and dynamical properties by focusing on the positions and motion of the cells' centres of mass.
\label{fig:2}
}
\end{figure*}

We assess the role of adhesion, motility, density and \textsc{cil} by evaluating the steady-state structural and dynamical properties in the $\tilde f$--$\phi_{\rm eff}$ plane.
$\tilde f = f^{\rm sp}/f_{\rm ad}$ plays the role of the temperature-to-adhesion energy ratio relevant to thermal systems.
For $\tilde f < 1$, two isolated adhering particles cannot detach through their self-propelling forces.
We investigate these diagrams in the absence and with \textsc{cil}, fixing $\tau_{\textsc{cil}}D_r = 10^2$ and $\tau_{\textsc{cil}}D_r = 10^{-4}$, and report them in Fig.~\ref{fig:2}(a) and (b).

Decreasing propelling forces induces a gas-liquid phase separation at low densities and the arrest of the dynamics in a glassy phase at high densities.
We identify the coexisting volume fractions with those at which the bimodal distribution of the local volume fraction $P(\phi_{\rm loc})$ peaks, as Fig.~\ref{fig:2}(e) illustrates.
The local volume fraction is measured on circular regions of radius $2.5d_0$, as detailed in SI Fig.~S4.

The \textsc{no-cil} diagram parallels the temperature-density phase diagram of sticky colloidal particles~\cite{Royall2018}, but for the absence of a re-entrant glass transition.
This finding suggests that with \textsc{no-cil} cells are in-near equilibrium, owing to their small persistent length. 
To confirm this result, we show in SI Fig. S5 that equilibrium simulations of this model lead to a phase diagram very close to the \textsc{no-cil} one.

\textsc{cil} drastically influences the phase diagram and the dynamics.
It reduces the maximum activity at which coexistence occurs and shrinks the width of the coexistence region, opposing density fluctuations.
It also suppresses density fluctuations in the homogeneous phase~\cite{Smeets2016a}.
Fig.~\ref{fig:2}(f) demonstrates that with \textsc{no-cil}, the fluctuations vanish at low density and confluence, and have a maximum at intermediate densities. 
Conversely, with \textsc{cil} the fluctuations are notably dampened, decreasing consistently with density and vanishing for $\phi_{\rm eff} \lesssim 1$.
This suppression of the density fluctuations does not result from drastic changes in the underlying ordering properties.
Indeed, the radial distribution function and the distribution of the local hexatic order parameter (see SI for detail) illustrated in Figs.~\ref{fig:2}(g) and (h) clarify that \textsc{cil} only promotes order at high densities.

The structural and dynamic changes induced by \textsc{cil} are related unexpectedly. 
The shrinkage of the coexistence region may suggest \textsc{cil} promotes particle motion, effectively making the tissue `hotter'. 
Instead, we find that \textsc{cil} commonly inhibits motion and promotes a glass-like dynamics, as demonstrated by the mean square displacement and the non-Gaussian parameter of the \textsc{cil} and \textsc{no-cil} dynamics illustrated in Fig.~\ref{fig:2}(i) and (j).
However, \textsc{cil}'s influence on the dynamics depends on the density. 
Fig.~\ref{fig:2}(k) reveals that while the \textsc{no-cil} diffusivity consistently decreases with the density, the \textsc{cil} one does not. 
In our model, the dynamics speeds up upon compression above confluency as the absence of free space makes \textsc{cil} irrelevant, and the \textsc{cil} and \textsc{no-cil} diffusivities converge. 
While a speedup of the dynamics upon compression (not overcrowding) has been observed in some experiments~\cite{Park2015, Mitchel2020}, this regime might be experimentally inaccessible as at high densities, cell extrusion processes set in~\cite{Marinari2012}.

We find that \textsc{cil} primarily inhibits particle motion by affecting the correlation time of cells' polarities.
To show that this is the case, we consider that the diffusivity scales as $D = \kappa v_0^2\tau_r$, $\kappa$ being a constant depending on the structure, $v_0$ the typical particle velocity, and $\tau_r$ the polarity correlation time.
\textsc{cil} does not significantly influence the characteristic particle velocity, as apparent from the collapse of the mean square displacement of the \textsc{cil} and \textsc{no-cil} dynamics in the ballistic regime, where $\langle r^2 \rangle = (v_0t)^2$.
Conversely, \textsc{cil} affects both the structure, and hence $\kappa$, and $\tau_r$ by promoting the repolarization of a cell towards the free space probed by its lamellipodium.
If \textsc{cil} primarily influences the diffusivity by affecting the polarity correlation time, which equals $D_r^{-1}$ in its absence, then \textsc{cil} diffusivity should scale as $D_{\rm predicted}=D_{\textsc{no-cil}}\tau_r D_r$. 
To investigate this hypothesis, we evaluate the repolarization timescale $\tau_r$ from the exponential decay of the correlation function of the polarities $C_p(t) = \langle {\bf p}(t) {\bf p}(0) \rangle = e^{-t/\tau_r}$ illustrated in Fig.~\ref{fig:2}(l).
Fig.~\ref{fig:2}(k) show that $D_{\rm predicted}$ approximate $D_{\rm cil}$, indicating that much of $\textsc{cil}$'s influence on the macroscopic tissue properties results from the change in the repolarization timescale.
The approximation is worse close to confluency, where $\textsc{cil}$ more strongly influences the structural properties.

\begin{figure*}[t!]
\centering
\includegraphics[width=\textwidth]{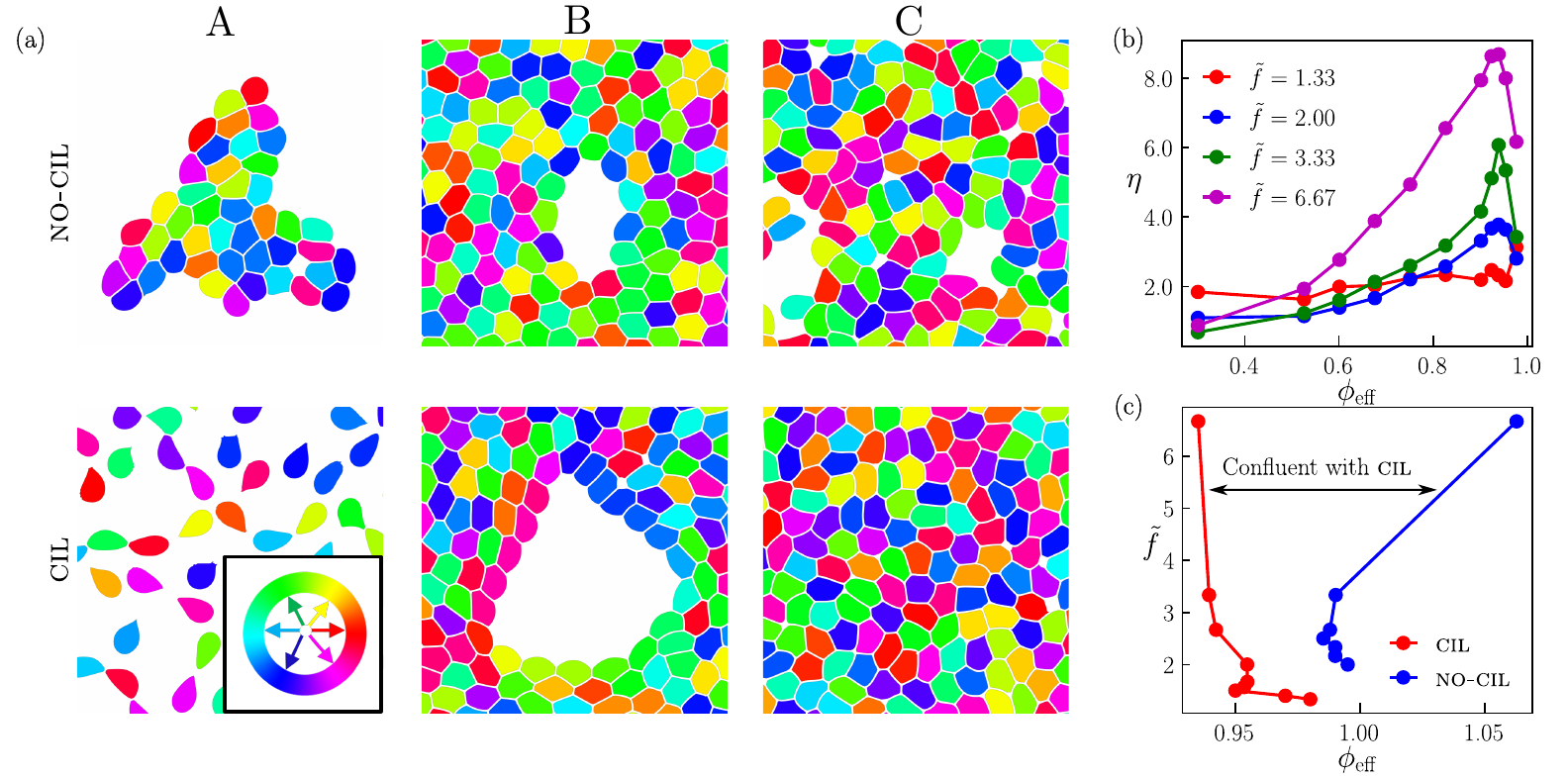}
\caption{
(a) The snapshots show a fraction of the system at state points A, B, and C in Fig.~\ref{fig:2}(a) and (b) for both \textsc{no-cil} and \textsc{cil} dynamics. Particles are colored based on their polarity, as indicated by the colour wheel in the bottom left corner. 
\textsc{cil} places clusters under tensions, possibly breaking them (A); exerts tensile forces at the rim of cavities (B), which may lead to wound healing (C).
(b) The relative change in effective area fraction induced by \textsc{cil}, $\eta=\frac{\phi_\textsc{cil}-\phi_\textsc{no-cil}}{\phi_\textsc{no-cil}}$, increases with motile strength $\tilde f$, but decreases above confluency.
(c) Confluence lines on the $\phi_{\rm eff}-\tilde{f}$ diagram illustrate that \textsc{cil}, by swelling particles and placing the system under tension, reduces the effective area fraction needed to reach confluency at a fixed motile strength $\tilde{f}$.
\label{fig:3}
}
\end{figure*}

Images of the system with particles coloured according to their polarities vividly illustrate the mechanisms through which \textsc{cil} influences structure and dynamics (see also SI Fig. S6).
In Fig.~\ref{fig:3}, we provide these images focussing on the state points marked with A, B and C in Fig.~\ref{fig:2}(a) and (b).
At low density and low activity (point A), the \textsc{no-cil} system is phase-separated and comprises several small clusters, with the polarity of each cell being random. 
In the presence of \textsc{cil}, the polarities of cells on the clusters' surface orient toward the free space, exerting tensile forces that break the clusters. 
This mechanism suppresses density fluctuations and drives the system into the homogeneous phase.
In the high-density and small-activity point B, which lies within the coexistence region, the system comprises a large confluent network including many small gas cavities with $\textsc{no-cil}$, and a single large one with $\textsc{cil}$.
With \textsc{cil}, cells on the rim have polarities directed toward the centre and exert tensile forces, which deform the tissue and increase the overall area fraction.
Since the tensile forces are small, the increase in area fraction is not large enough for the system to reach confluency.
In C, we consider the same density as B, but larger self-propelling forces. 
In this case, with \textsc{cil}, the change in area is large enough to close cavities and bring the system to confluence, while with \textsc{no-cil} the system reaches a non-confluent homogeneous state. 
The change in area fraction is visually apparent in Fig.~\ref{fig:2}d, where we show configurations corresponding to point C with the particle colour-coded according to the actual shape index $q_i=P_i/\sqrt{A_i}$.
Therefore, with \textsc{cil}, the closure of cavities moving from B to C corresponds to a \textsc{cil}-facilitated motility-induced suppression of density fluctuations: wound healing.

The relative area fraction $\eta=\frac{\phi-\phi_{\rm eff}}{\phi_{\rm eff}}$ change, which equals the relative change in the average particle area,
quantifies \textsc{cil}'s ability to deform the particles. 
Fig.~\ref{fig:3}(b) shows that, below confluence $\phi \simeq 1$, $\eta$ increases with $\phi_{\rm eff}$ and with $\tilde f$, which sets the magnitude of the tension force.
Henceforth, tissues reach confluency at smaller densities in the presence of \textsc{cil}, as we demonstrate by comparing the \textsc{cil} and \textsc{no-cil} confluency lines in Fig.~\ref{fig:3}(c).
In the region enclosed by the two lines, tissues are only confluent with \textsc{cil}. 

An ambivalent cell tendency shapes the \textsc{cil} phase diagram:
Cell adhesion promotes aggregation and, hence, density fluctuations;
Conversely, \textsc{cil} counteracts density fluctuations by repolarising cells away from contacts,
swelling the cells or breaking the adhesive junctions.
The interplay between \textsc{cil} and adhesion is critical for attaining confluency under tension.

\section*{Wound healing and tensile force chains}
\begin{figure*}[t!]
\centering
\includegraphics[width=0.9\textwidth]{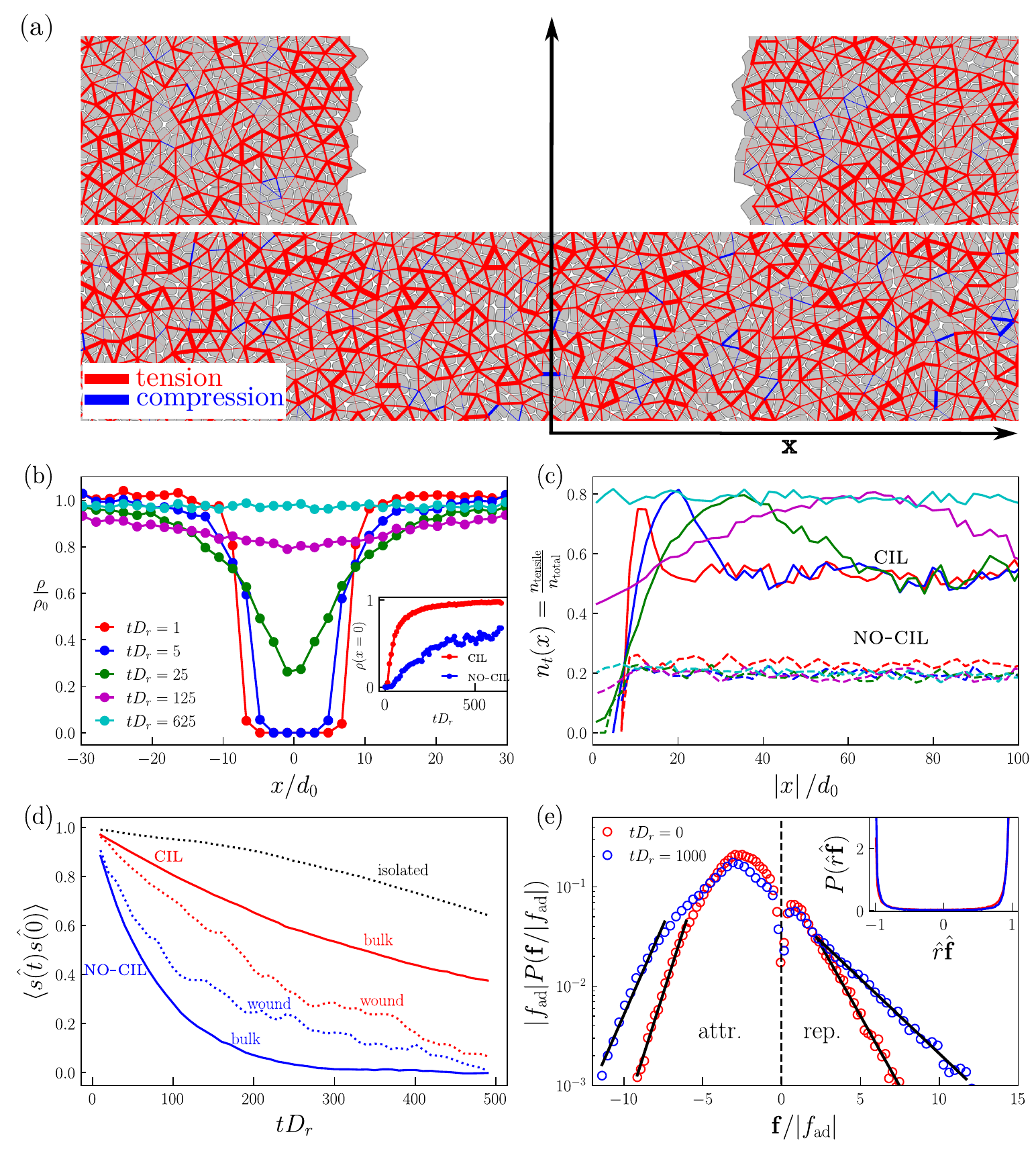}
\caption{
(a) Wound healing assay experiment with \textsc{cil} dynamics: initial and final configurations.
In the horizontal direction, we illustrate a small fraction of the system.
Spatial dependence (b) of the monomer density profile normalised to its initial average value $\rho_0$, in the presence of $\textsc{cil}$ (see SI for $\textsc{no-cil}$), and (c) of the fraction of monomers under tensile force, with and without $\textsc{cil}$.
Lines refer to diverse times, with $tD_r=400$ corresponding to approximately $8$ days.
The inset of (b) illustrates the time dependence of the density at the wound centre.
(d) The decay of the correlation function of the versor $\hat s$ pointing from a cell centre of mass to a material point (monomer) on its surface demonstrates cell rotational motion, in different conditions.
%With $\textsc{cil}$, rotations are faster close to the wound that in the bulk, while the converse occurs in its absence.
(e, inset) the probability distribution of $\hat {\bf r}\cdot \hat {\bf f}$, with ${\bf r}$ the distance between the centres of masses of two interacting cells, and ${\bf f}$ their net interaction force, demonstrates that interaction forces are essentially central.
(e) Probability distribution of the net magnitude of the net intercellular interaction force, $|{\bf f}|/f_{\rm ad}$, before scratching and in the final healed state. Positive and negative values correspond to tensile and compressive forces.
In (b) and (c), data are coarse-grained over bins of width two cell diameters.
In (c), we also average the $\pm x$ symmetric directions.
All data are averaged over 5 numerical experiments. 
}
\label{fig:healing}
\end{figure*}

To elucidate how \textsc{cil} influences the dynamics and the mechanics of the healing process, we investigate a system of $N = 10,000$ cells enclosed in a strip with a width of 15 cell diameters in a scratch-assay configuration. 
Fig.~\ref{fig:healing}(a) shows a small fraction of the investigated system. 
Healing occurs as cells at the leading edge, along with those a few rows behind it~\cite{Trepat2009}, polarize and crawl toward the wound. 
Here, we focus on \textsc{cil}'s influence and the strength and heterogeneity of the intercellular forces, which are of difficult experimental measure. 
%We exploit our model to elucidate the mechanical forces acting during the healing process and the influence of \textsc{cil}. 
%To achieve this, we investigate healing in a system of $N = 10,000$ cells enclosed in a strip with a width of 15 typical cell diameters.

We fix an active force $\tilde{f} = 6.67$, which is larger than that at the `critical point' (see Fig.~\ref{fig:2}(a),(b)), and set the initial density (post-scratch) between the \textsc{cil} and \textsc{no-cil} confluency lines (see Fig.~\ref{fig:3}(c)). 
These parameters ensure the system reaches a homogeneous configuration confluent with \textsc{cil}, but not without (SI Fig.~S7).
At smaller $\tilde{f}$, the system heals with \textsc{cil} while it remains wounded, i.e., phase-separated, in its absence, as we demonstrate in SI Fig.~S8.
%In the SI, we show analogous results for smaller $\tilde{f}$, where the \textsc{cil} system heals, but the \textsc{no-CIL} system remains wounded, i.e., phase-separated.

Fig.~\ref{fig:healing}(b) illustrates the evolution of the monomer density profile $\rho(x,t)$, coarse-grained over bins with a width of approximately 2 cell diameters and normalized by the average density before the scratch. 
As healing progresses, the cell density increases in the wounded region while slightly decreasing farther away. 
At long times, the density saturates to a value slightly smaller than the initial one. 
Confluency, as demonstrated in the bottom panel of Fig.~\ref{fig:healing}(a), is achieved due to particle swelling.
In the SI, we present the evolution of the density profile without \textsc{cil}, where cells do not swell, and the system settles into a non-confluent steady state. 
The inset of Fig.~\ref{fig:healing}(b) shows the time dependence of the density at the wound center, demonstrating that \textsc{cil} shortens the time required to reach the asymptotic steady state. 
Thus, while \textsc{cil} slows down motion in a steady state (as seen in Fig.~\ref{fig:2}(k)), it accelerates the healing dynamics.

The migrating cells remain generally anchored to the surrounding tissue, generating a tension wave, which we reveal by analyzing the spatial dependence of the fraction of monomers under tension, $n_t(x)$. 
We consider a monomer $i$ at a distance ${\bf r}_i$ from the cell's centre of mass under tension if the force ${\bf f}_i$ it experiences from interactions with monomers of other cells satisfies ${\bf f}_i \cdot {\bf r}_i > 0$. 
Fig.~\ref{fig:healing}(d) shows that $n_t(x)$ peaks at a distance $x_{\rm tensile}(t)$ from the wound center and asymptotically approaches the value typical of the initial unscratched tissue at greater distances.
The position of this maximum, $x_{\rm tensile}(t)$, increases over time, indicating the propagation of a tensile wave. At early times, the velocity of this wave is approximately $10 \unit{\um/h}$. 
A tension wave propagation is consistent with previous experimental findings~\cite{Trepat2012}.
These experiments also reported that inhibiting cell-cell adhesion disrupts monolayer integrity and, consequently, the tensile wave. 
Consistently, we find that \textsc{cil}-suppression, which also disrupts confluence, prevents the propagation of the tensile wave. 
Indeed, Fig.~\ref{fig:healing}(d) shows that without \textsc{cil}, $n_t(x) \approx 0.2$ remains nearly constant across space and time.

Cells are extended objects that, besides migrating, can also individually and collectively rotate.
However, exposing individual cell rotations is challenging because it requires dynamic information at the subcellular scale, and thus has remained largely unexplored.
To demonstrate cell rotations, we associate with each cell a versor $\hat{\bf s}(t)$ pointing from the cell's centre of mass to a material point on its surface, representing a specific cell monomer.
Fig.~\ref{fig:healing}(c) shows the evolution of the correlation function $\langle \hat{\bf s}(t) \hat{\bf s}(0) \rangle$, averaged over cells near the wound edge ($|x| < 20$) and farther away, comparing dynamics under \textsc{cil} and \textsc{no-cil} conditions. 
%As a reference, we also illustrate the correlation function for isolated cells.
Interestingly, \textsc{cil} suppresses the rotational motion, and it does so more in the bulk than close to the leading edge.

What makes a cell rotate?
In our model, the active force on a monomer is aligned to the versor connecting it to the cell's centre of mass (see Fig.~\ref{fig:model}a) and cannot generate any torque. 
Cell-cell interaction forces may generate torques. 
However, we find the net cell-cell interaction force $\bf{f}$ between two particles, which is the sum of forces between monomers of the cells, is typically aligned with the line connecting their centres of mass, $\bf{r}$, as illustrated in the inset of Fig.~\ref{fig:healing}(d), indicating that these forces do not induce torques.
Instead, torques are mainly induced by the non-central frictional forces with the substrate.
Cell deformation is essential for producing a net torque; without deformation, the torques exerted by individual monomers would cancel out.
Indeed, we show in Fig.~\ref{fig:healing}(d) that the correlation function for isolated cells, whose shape fluctuations are minimal, decays over an extremely log time.
At the considered densities, cells are more regular with than without \textsc{cil}, as apparent by comparing Fig.~\ref{fig:2}(c) and (d) and demonstrated by the structural characterizations in Fig.~\ref{fig:2}(g) and (h).
This regularity explains why \textsc{cil} suppresses rotational motion and why particles close to the leading edge, which are less ordered, rotate more than particles in the bulk.

The alignment between the interaction force and the line connecting the centres of two interacting cells suggests a force network description of intercellular forces~\cite{Liu1995}.
This framework is unexplored in tissue contexts and, more broadly, in systems of deformable particles. 
Fig.~\ref{fig:healing}(e) illustrates the distribution of the intercellular force magnitude, normalized by the minimal active force required to detach two adherent cells, $f_{\rm ad}$, both just before wound creation and in the asymptotic final state. 
The distribution exhibits an exponential decay at large repulsive and attractive forces, resembling the behaviour observed in compressed wet granular particles, where tensile forces are mediated by liquid bridges~\cite{Radjai2010}.
We highlight that most forces are attractive thanks to \textsc{cil}. 
In its absence, particles persistently push on each other, and interaction forces are primarily repulsive (SI Fig. S7).

The exponential decay of the forces reveals strong force heterogeneities, i.e., the coexistence of small and much larger compressive and tensile forces, which are larger in the final healed state.
The force heterogeneity is apparent in Fig.~\ref{fig:healing}(a), where we represent the intercellular forces as segments connecting the centres of interacting cells.
The width of each segment is proportional to the magnitude of the interaction force, and the colours distinguish tensile (red) from compressive (blue) forces.
In systems of stiff, repulsive particles, the large forces organise in one-dimensional linear structures or force chains.
Here, we observe the emergence of one-dimensional chains of large attractive forces. 
However, these chains may not be linear, as we clarify in SI Fig. S9.
Since cells can actively sense and respond to their interaction forces, the observed heterogeneity in the force network may have significant biological implications that warrant further investigation.

\section*{Conclusion and Discussion}
We have introduced a mechanical model for isolated two-dimensional epithelial cells that offers enhanced spatial resolution while remaining computationally efficient.
The model provides detailed descriptions of lamellipodium-induced dynamics, adhesion mechanisms, contact inhibition of locomotion, and forces at play, which are crucial to rationalising how their interplay regulates tissue properties.

Without \textsc{cil}, tissues behave as sticky colloidal particles: motility alone does not introduce a distinctive phenomenology due to the small cells' persistence length.
\textsc{cil} emerges as a microscopic mechanism through which tissues suppress density fluctuations.
In a steady state, where the density fluctuations are slight, \textsc{cil} slowdowns the dynamics by inhibiting the structural rearrangements driving diffusion, which entail localised fluctuations in the density.
Conversely, \textsc{cil} speeds up the relaxation dynamics when the initial state has large density fluctuations, as in the case of a wound.
%, and its interplay with adhesion enables tissues to set under tension spontaneously. 
%\textsc{cil}'s suppression of density fluctuations has contrasting effects on the dynamics.
%In a steady state, where the density fluctuations are slight, \textsc{cil} slowdowns the dynamics as it inhibits the structural rearrangements %driving diffusion, given that these entail localised density changes.
%Conversely, \textsc{cil} speeds up the relaxation dynamics when the initial state has large density fluctuations, as in the case of a wound.
\textsc{cil} influence on the dynamics enables the emergence of tensile force chains crucial for tissues' mechanical stability. 
Our work highlights \textsc{cil}'s importance in maintaining tissue integrity and promoting effective wound healing, providing a deeper understanding of the mechanical and dynamic interactions in epithelial tissues.

The mechanisms influencing tissue properties are numerous and often debated, making it challenging to investigate them all.
For instance, there is indirect evidence suggesting the existence of polarity alignment mechanisms driven by stress~\cite{Tambe2011,Trepat2011}, flow or cell-cell interactions~\cite{Alert2020}; there could be a coupling between self-propulsion strengths and density~\cite{Garcia2015}, cell-junction mediated frictional forces might be relevant, and on longer time, cell mechanical properties might exhibit  plasticity~\cite{Wyatt2016}.
Due to its enhanced spatial resolution, our model can certainly be used to investigate the relevance of these effects.
It can be similarly used to investigate tissue and cell cluster motion~\cite{Doxzen2013,Copenhagen2018,Ladoux2017a}, as well as tissue mechanical properties, e.g., fracture under tensions~\cite{Harris2012}.
%, as we illustrate in the SI.
%We hope the research community will help develop our model into one that accounts for cellular behaviour in various contexts.

%\mpc{aaaa}
%While healing is facilitated by cell reproduction, and reproduction is critical for the healing of large wounds, the invasion process is %largely insensitive to it~\cite{Tlili2018,Gauquelin2019}.
%In this context, our model offers the opportunity to clarify \textsc{cil} role in the dynamics and the mechanics of the healing properties.

\backmatter

\bmhead{Supplementary information}
This article has accompanying supplementary files. 

\bmhead{Acknowledgments}
We sincerely thank G. Charras and N. Gov for their valuable comments and feedback, which have helped improve this work.
This research was supported by the Singapore Ministry of Education through grants
MOE-T2EP50221-0016 and RG152/23

\section*{Declarations}
The authors declare no competing financial interests. 
Supplementary information accompanies this paper on www.nature.com/naturephysics. 
Our numerical code, NexTissUe, which is an extension of the LAMMPS package, is available at \url{https://github.com/anshtg20/NextTissUe}

Reprints and permissions
information is available online at www.nature.com/reprints. Correspondence and
requests for materials should be addressed to M.P.C.

\bibliography{tissue,sn-bibliography}% common bib file
%% if required, the content of .bbl file can be included here once bbl is generated
%%\input sn-article.bbl

\end{document}

%\title[Article Title]{Article Title}
\title[Tissue]{Supplementary Information for `Epithelial Tissues from the Bottom-Up: Contact Inhibition, Wound Healing, and Force Networks'}

%%=============================================================%%
%% Prefix	-> \pfx{Dr}
%% GivenName	-> \fnm{Joergen W.}
%% Particle	-> \spfx{van der} -> surname prefix
%% FamilyName	-> \sur{Ploeg}
%% Suffix	-> \sfx{IV}
%% NatureName	-> \tanm{Poet Laureate} -> Title after name
%% Degrees	-> \dgr{MSc, PhD}
%% \author*[1,2]{\pfx{Dr} \fnm{Joergen W.} \spfx{van der} \sur{Ploeg} \sfx{IV} \tanm{Poet Laureate} 
%%                 \dgr{MSc, PhD}}\email{iauthor@gmail.com}
%%=============================================================%%

\author[1]{\fnm{Anshuman} \sur{Pasupalak}}
\author[1]{\sur{Zeng} \fnm{Wu}}
\author[1,2]{\fnm{Massimo} \sur{Pica Ciamarra}}\email{massimo@ntu.edu.sg}

\affil[1]{\orgdiv{Division of Physics and Applied Physics, School of Physical and Mathematical Sciences}, \orgname{Nanyang Technological University}, \orgaddress{\street{21 Nanyang Link}, \postcode{637371}, \country{Singapore}}}
\affil[2]{\orgdiv{Consiglio Nazionale delle Ricerce}, \orgname{CNR-SPIN}, \orgaddress{\city{Napoli}, \postcode{I-80126}, \country{Italy}}}

%\keywords{keyword1, Keyword2, Keyword3, Keyword4}
%%\pacs[JEL Classification]{D8, H51}
%%\pacs[MSC Classification]{35A01, 65L10, 65L12, 65L20, 65L70}

\maketitle

\tableofcontents

\newpage
\section{Monomer-monomer interaction potential}
\begin{figure}[h!!]
\centering
\includegraphics[width=0.7\textwidth]{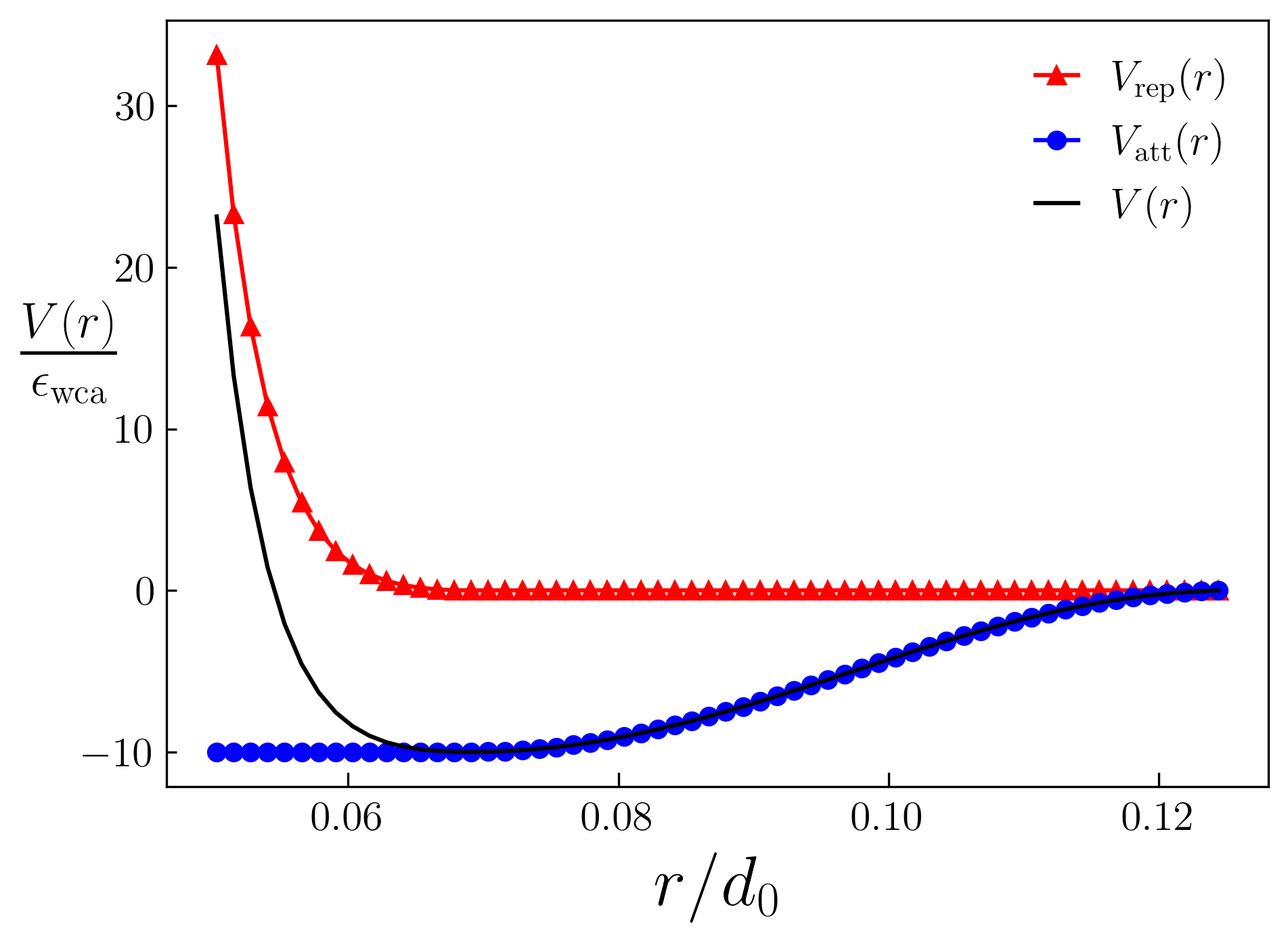}
\caption{
Interaction potential between monomers of distinct cells. 
Distances are normalised by the typical cell linear size.
\label{fig:interaction}
}
\end{figure}

Monomers making up distinct cells interact via the interaction potential illustrated in Fig.~\ref{fig:interaction}.
This potential is the sum of purely repulsive and attractive ones, which models steric repulsion and short-range attraction.
\begin{equation}
V(r) = V_{\rm rep}(r) + V_{\rm att}(r)
\end{equation}
\begin{equation}
	\label{pairwise_pot}
	\begin{split} V_{\rm rep}(r) =
		\begin{cases}
			\epsilon_{\rm wca} \left[ \left(\frac{\sigma}{r}\right)^{12} - 2\left(\frac{\sigma}{r}\right)^6 + 1\right] & \quad r < \sigma \\
			0& \quad r \geq \sigma
		\end{cases}
	\end{split}
\end{equation}
\begin{equation}
	\label{pairwise_pot}
	\begin{split} V_{\rm att}(r) =
		\begin{cases}
  	     -\epsilon_{\rm a} &\quad r \leq \sigma \\
			-\epsilon_{\rm a} \cos\left(\frac{\pi\left(r - \sigma\right)}{2\left(r_c - \sigma\right)}\right)^2&\quad \sigma \leq r < r_c \\
			0& \quad r \geq r_c
		\end{cases}
	\end{split}
\end{equation}
We fix $\epsilon_{\rm a} = 10 \epsilon_{\rm wca} $. 
In our simulations, each cell has $N_m=50$ monomers. 
In general, $\epsilon_{\rm wca}$ needs to scale as $1/N_m$ for the cell-cell adhesion energy to be $N_m$ independent.

The monomer diameter and the width of the attractive well have a length corresponding to a minute fraction of a cell diameter, $d_0$. 

\newpage
\section{Free space probed by the lamellipodium}
\begin{figure}[h!]
\centering
\includegraphics[width=0.5\textwidth]{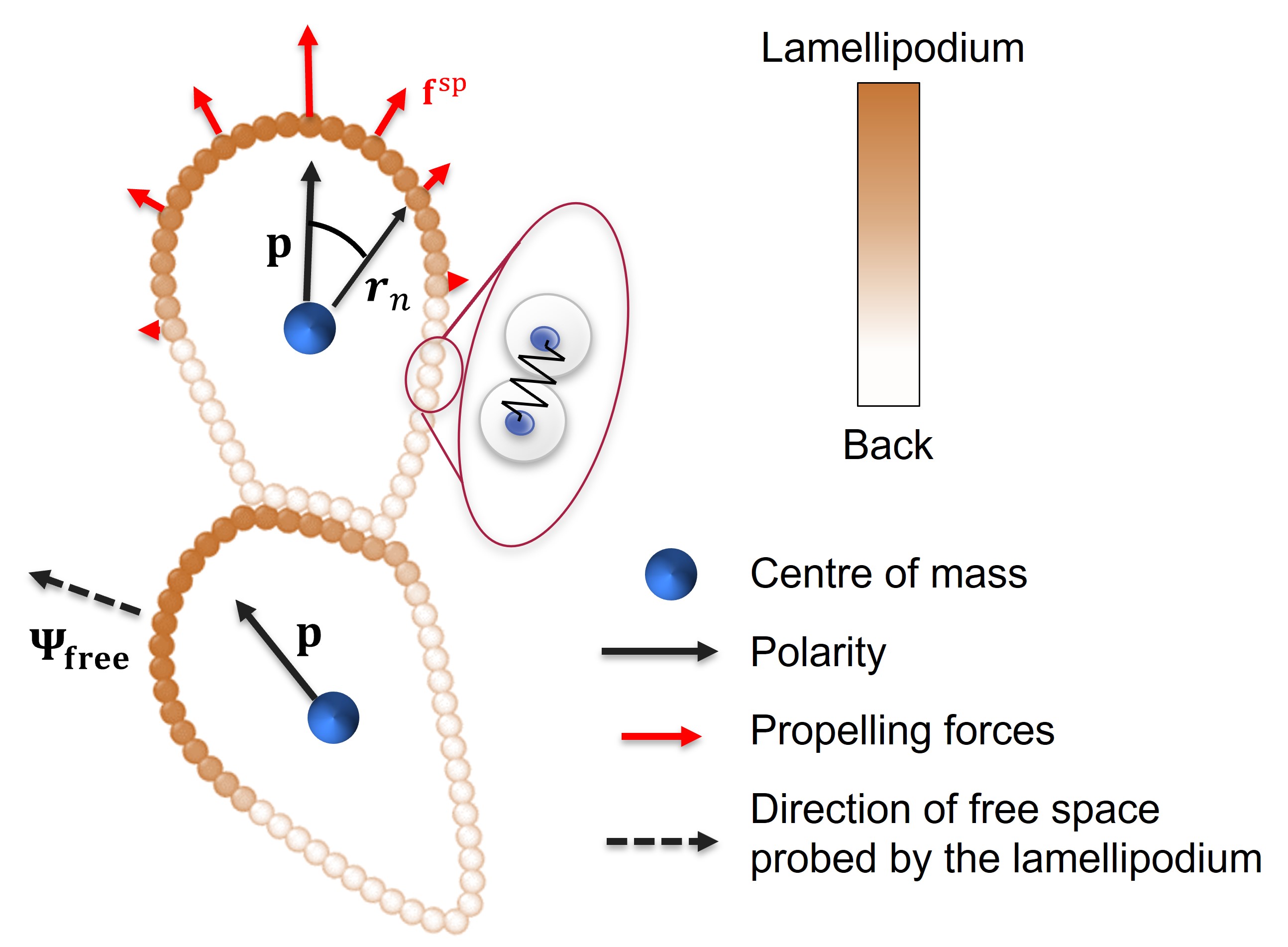}
\caption{As in Fig.1 in the main text.
\label{fig:free}
}
\end{figure}

We model \textsc{cil} as an alignment interaction between a cell's polarity and the direction of the free space proper by its lamellipodium, ${\bf \Psi}_{\rm free}$. 
This direction is illustrated in Fig.~\ref{fig:free}, for the bottom cell.

We define the versor ${\bf \Psi}_{\rm free}$ as a weighted average of the distance ${\bf r}$ of the cells' monomers from the cell's centre of mass. 
Monomers that are not free, as they interact with monomers of other cells ($|F_{\rm int}| \neq 0$), have weight zero.
The other monomers' weight depends on the scalar product $\hat{\bf r} \cdot \hat{\bf p}$. 
This dependence ensures the monomers at the front, which define the lamellipodium, influence ${\bf \Psi}_{\rm free}$ more.
Specifically, we define
\begin{equation}
    {\bf \Psi}_{\rm free} = A \sum_{|F_{\rm int}|=0} \frac{(1+\hat{\bf r} \cdot \hat{\bf p})}{2}  \hat{r},
\end{equation}
with $A$ a normalisation constant.

${\bf \Psi}_{\rm free}$ is not defined for particles in confluent tissue, or, more generally, when no monomer is free from interactions. 
In that case, the sum reduces to the null vector $\vec{0}$.
The pathological behaviour is naturally resolved as \textsc{cil} has a strength that depends on the fraction $\beta$ of free monomers. 
When ${\bf \Psi}_{\rm free}=\vec{0}$, $\beta = 0$ and \textsc{cil} plays no role.

\newpage
\section{Measure of the force scale, $\tilde f$}
In our model, each cell monomer is assigned a scalar $0 \leq \Lambda = \max({\bf p \cdot r},0) \leq 1$ that measures its contribution to the lamellipodium. Here, ${\bf p}$ is the cell polarity and ${\bf r}$ is the monomer position, ${\bf r = 0}$ being that of the cell centre of mass.
A monomer-dependent active force $f_0\Lambda \hat {\bf r}$ drives cell motion, and additional viscous force proportional to the velocity $-\gamma \dot {\bf r}$ ensures the overdamped regime.

\begin{figure}[h!]
\centering
\includegraphics[width=1\textwidth]{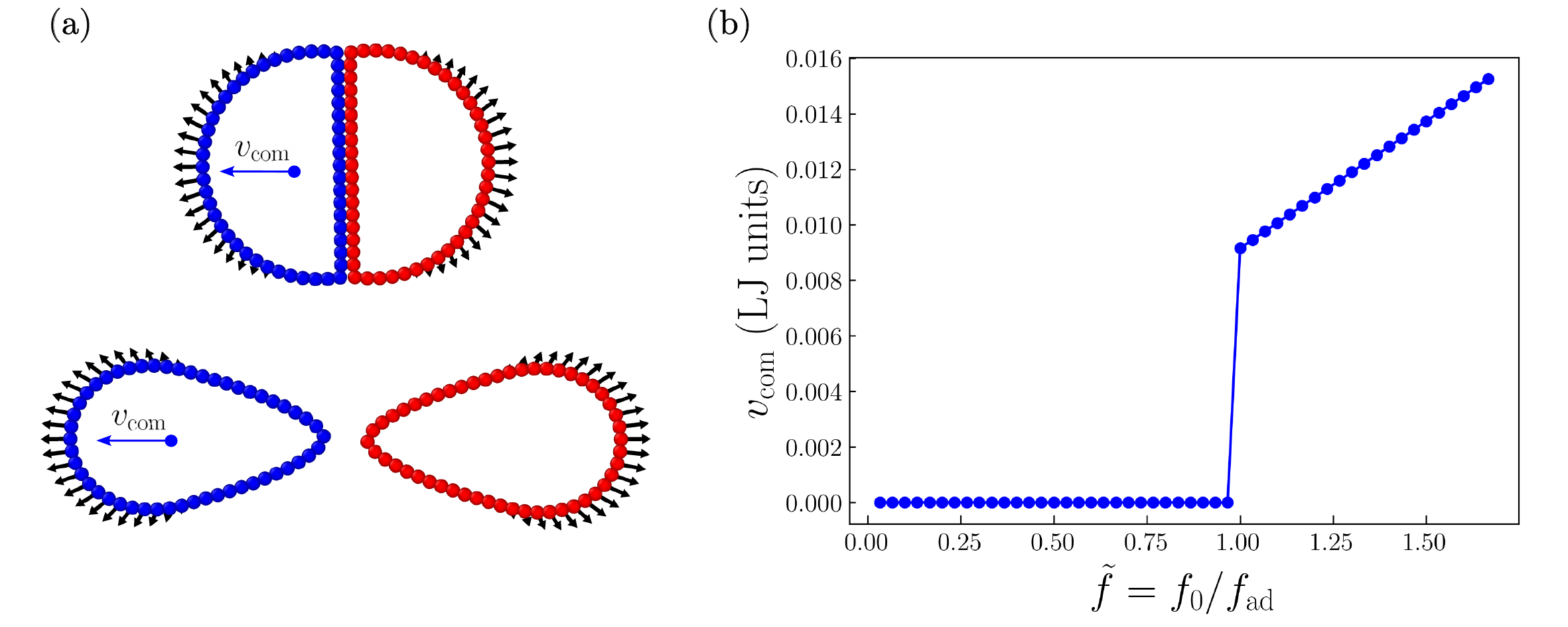}
\caption{(a) Geometric set-up used to investigate the detachment of two adherent cells on increasing the self-propelling force.
(b) the velocity of the centre of mass of each cell is zero before yield, and then increases linearly with the active force. 
\label{fig:forcescale}
}
\end{figure}

We set the scale of $f_0$ via an ideal numerical experiment involving two infinitely persistent cells ($D_r = 0$). 
The cells are initially adherent, with $f_0 = 0$, as in Fig.~\ref{fig:forcescale}(a).
Starting from this configuration, we quasistatically increase $f_0$ and measure the velocity of the centre of mass of each cell at each $ f_0$.
Due to the systems' symmetry, the two cells' velocities are equal and opposite.
For small $f_0$, adhesion dominates, and in a steady state, the cells do not move.
When $f_0$ overcomes a critical yielding threshold $f_{\rm ad}$, the two cells detach and move with a finite velocity proportional to $f_0$, as Fig.~\ref{fig:forcescale}(b) illustrates.
We use the yielding force as a force unit measure and report in the main text forces in terms of $\tilde f = f_0/f_{\rm ad}$.
Experiments indicate that this yielding force is of the order of  $10^{-7}\;\unit{\newton}$ 
(Esfahani {\it et al.}, Proceedings of the National Academy of Sciences {\bf 118},  e2019347118 (2021)).

\newpage
\section{Area fraction}
\subsection*{Area fraction and effective area fraction}
The area fraction of the system as 
\begin{equation}
\phi = \rho \langle A\rangle
\end{equation}
with $\langle A\rangle$ the average particle area.

We compute the area of a cell as that of a polygon with $N_m=50$ vertices.
To account for the finite size of the monomers we assume the polygon vertices are located in ${\bf w}_i=({\bf r}_i+\sigma/2)\hat {\bf r}_i$, where ${\bf r_i}$ is the position of monomer $i$ if ${\bf r}={\bf 0}$ identities the position of the cell's centre of mass, and $\sigma$ is the length scale associated with the repulsive monomer-monomer interaction.

Since particle deswell when compressed, the area fraction saturates to one and does not inform on the degree of overcrowding at high densities.
To measure overcrowding, we resort to the effective area fraction
\begin{equation}
\phi_{\rm eff} = \rho A_1
\end{equation}
with $\rho$ the number density and $A_1$ the area of an isolated non-motile particle in its minimal energy configuration. 
%The area $A_1$ of an isolated cell in a circle of radius $w$, and the area of the polygon is well approximated by $A = \pi w^2$.

\subsection*{Local area fraction}
\begin{figure}[b!]
    \centering
    \includegraphics[width=0.45\textwidth]{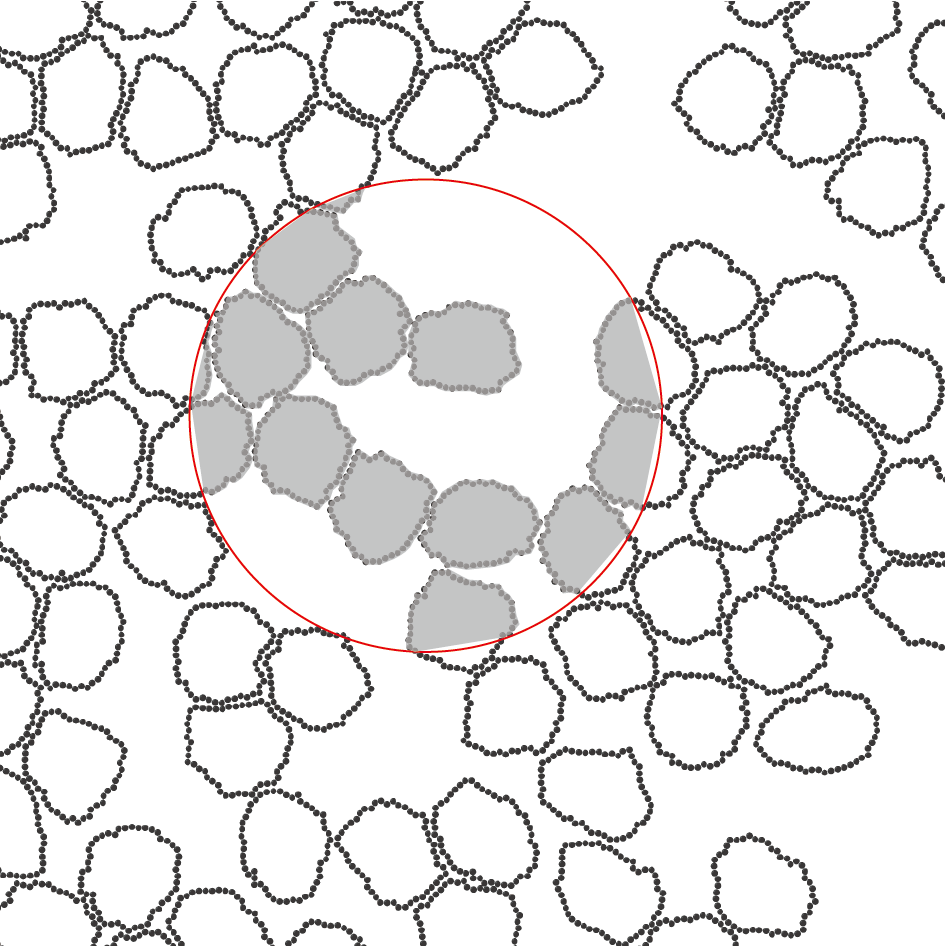}
    \caption{Schematic illustration of the approach used to measure the local area fraction $\phi_{\rm loc}$.}
    \label{fig:phi_loc}
\end{figure}
We measure the local area fraction in regions of radius $r_{\rm loc} = 2.5d_0$. 
To this end, we determine each cell's intersection area $A_k$ with the considered circle, by considering each cell as a polygon with vertices in ${\bf w}_i$.
The local volume fraction associated with the considered circle is $\phi_{\rm loc} = {\sum_k A_k}/{\pi r_{\rm loc}^2}$. 
See Fig.~\ref{fig:phi_loc} for an illustration.
To measure the distribution of the local volume fraction, we evaluate $\phi_{\rm loc}$ in 5000 circles with randomly selected centres.

\subsection*{Confluency}
The area fraction of the system coincides with the effective one at low density, while it saturates at $\phi_{\rm conf} \simeq 0.985$ at high density.  
We consider the system as confluent when the area fraction reaches this threshold.

\newpage
\section{Equilibrium phase diagram}
We have investigated our model's structural and dynamical in thermal equilibrium, to provide a reference scenario.
In this study, cells do not self-propel.
Conversely, they are driven by a Langevin dynamics, with the stochastic and damping forces acting on each cell centre of mass.

\begin{figure*}[h!]
\centering
\includegraphics[width=0.7\textwidth]{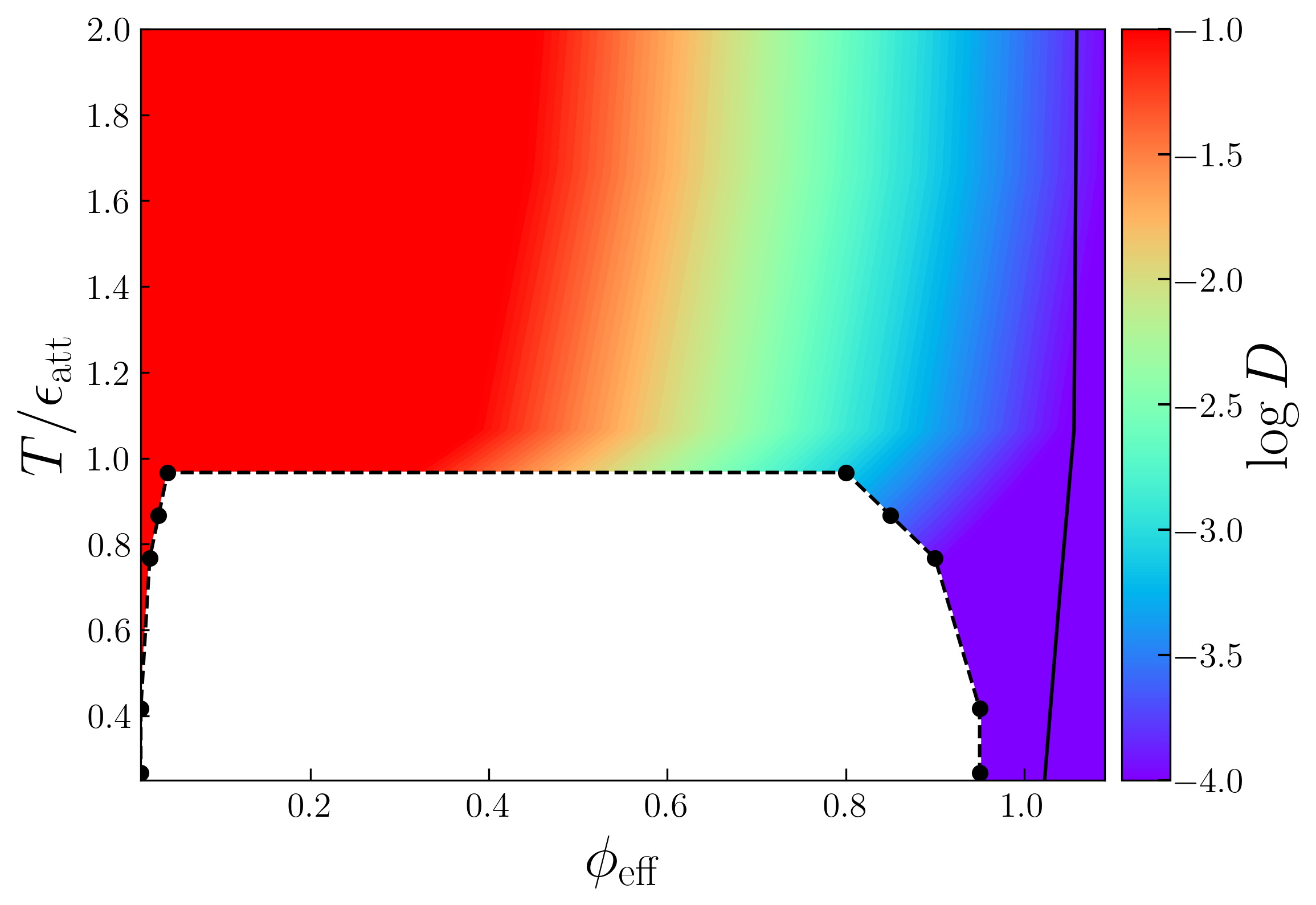}
\caption{
Thermal phase diagram.
\label{fig:T}
}
\end{figure*}

Fig.~\ref{fig:T} illustrates the resulting phase diagram, the colour code representing the cell diffusion coefficient.
This diagram parallels that observed in the active cell system with \textsc{no-cil} illustrated in Fig.~2(a) of the main text.

\newpage
\section{Snapshots of the system}
\begin{figure}[h!]
\centering
\includegraphics[width=0.8\textwidth]{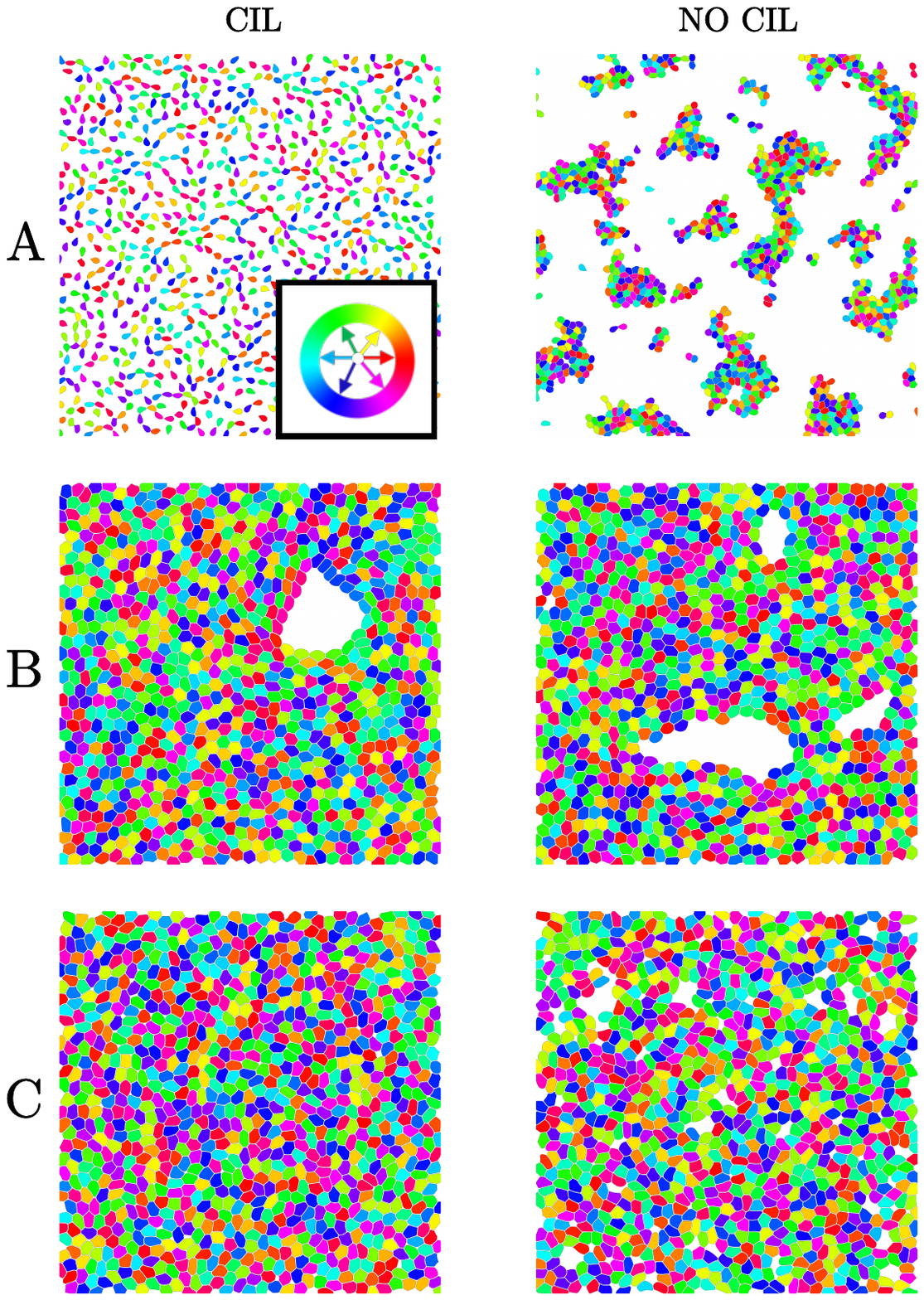}
\caption{Snapshots of the investigated system at the state points A, B, and C indicated in Fig. 2(a) and (b), from top to bottom,  with \textsc{cil} (left) and with \textsc{no-cil} (right). Particles are coloured according to their polarity.
 \label{snapshot}}
\end{figure}

\newpage

\section{Hexatic Order Parameter}
We investigate orientational order through the local hexatic order parameter
\begin{equation}
\Psi_{6,i} = \frac{1}{m}\sum_{j=1}^m e^{i6\theta_{ij}}
\end{equation}
Here, the sum is over all neighbours $j$ of cell $i$, and $\theta_{ij}$ is the angle between the radial vector joining cell centres $i$ and $j$ and a reference axis. 
We consider two cells as neighbours when their distance is smaller than that of the first peak of the radial distribution function.
The distribution of the magnitude of $|\Psi_{6,i}|$ illustrated in Fig.~2h in the main text proves the absence of hexatic ordering in the investigated range of parameters. Some hexatic order emerges at large densities in the presence of \textsc{cil}.

\newpage
\section{Wound healing: the role of \textsc{cil} and $\tilde f$}
In the main text, we have investigated the wound healing dynamical and mechanical properties at large $\tilde f$ and in the presence of \textsc{cil}. Here, we compare the \textsc{cil} and \textsc{no-cil} process, both at large and at smaller $\tilde f$

\subsection*{\textsc{cil} vs \textsc{no-cil} healing dynamics at large $\tilde f$}
\begin{figure*}[h!]
\centering
\includegraphics[width=0.90\textwidth]{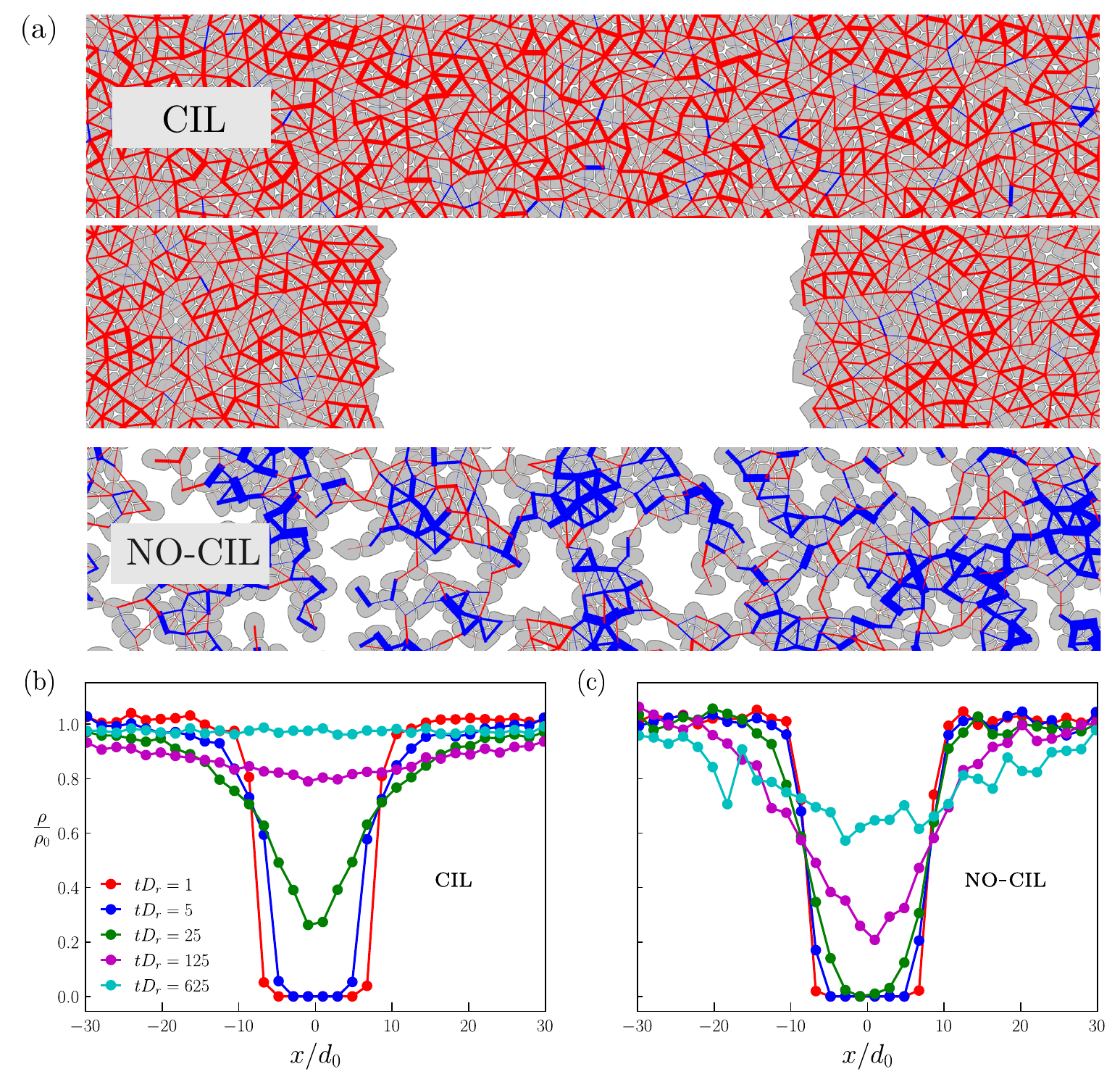}
\caption{
\textsc{cil} vs \textsc{no-col} healing dynamics at large $f$. With \textsc{cil}, the scratched configuration in the middle panel of (a) heals at long times, as visually demonstrated in the top panel and via the investigation of the monomer density profile (b). Conversely, with \textsc{no-cil}, the system evolves towards a non-confluent homogeneous state.
\label{fig:healing_large}
}
\end{figure*}

We consider the evolution of the scratched configuration illustrated in the middle panel of Fig.~\ref{fig:healing_large}(a).
At the considered $\tilde{f} = 6.67$ and initial density values the system is a state point following in the homogeneous region of the \textsc{cil} and the \textsc{no-cil} phase diagrams.
We detailed the evolution of this configuration in the presence of \textsc{cil} in the main text and recap here the final configuration in (a top panel) and the temporal evolution of the monomer density profile in (b).

With \textsc{no-cil}, the approach to the final steady state occurs on a time scale that is too large to investigate numerically. Indeed, the monomer density distribution does not flatten out over the longest times in our study. 
The bottom panel of (a) visually clarifies that the system evolves towards a non-confluent steady state where repulsive forces (blue) originating from the persistent cellular motion dominate.
In contrast, \textsc{cil} suppresses this persistence, as illustrated in Fig.~2(l), allowing adhesive forces (red) to dominate.

~\newpage~\newpage
\subsection*{\textsc{cil} vs \textsc{no-cil} healing dynamics at small $\tilde f$}
\begin{figure*}[h!]
\centering
\includegraphics[width=0.90\textwidth]{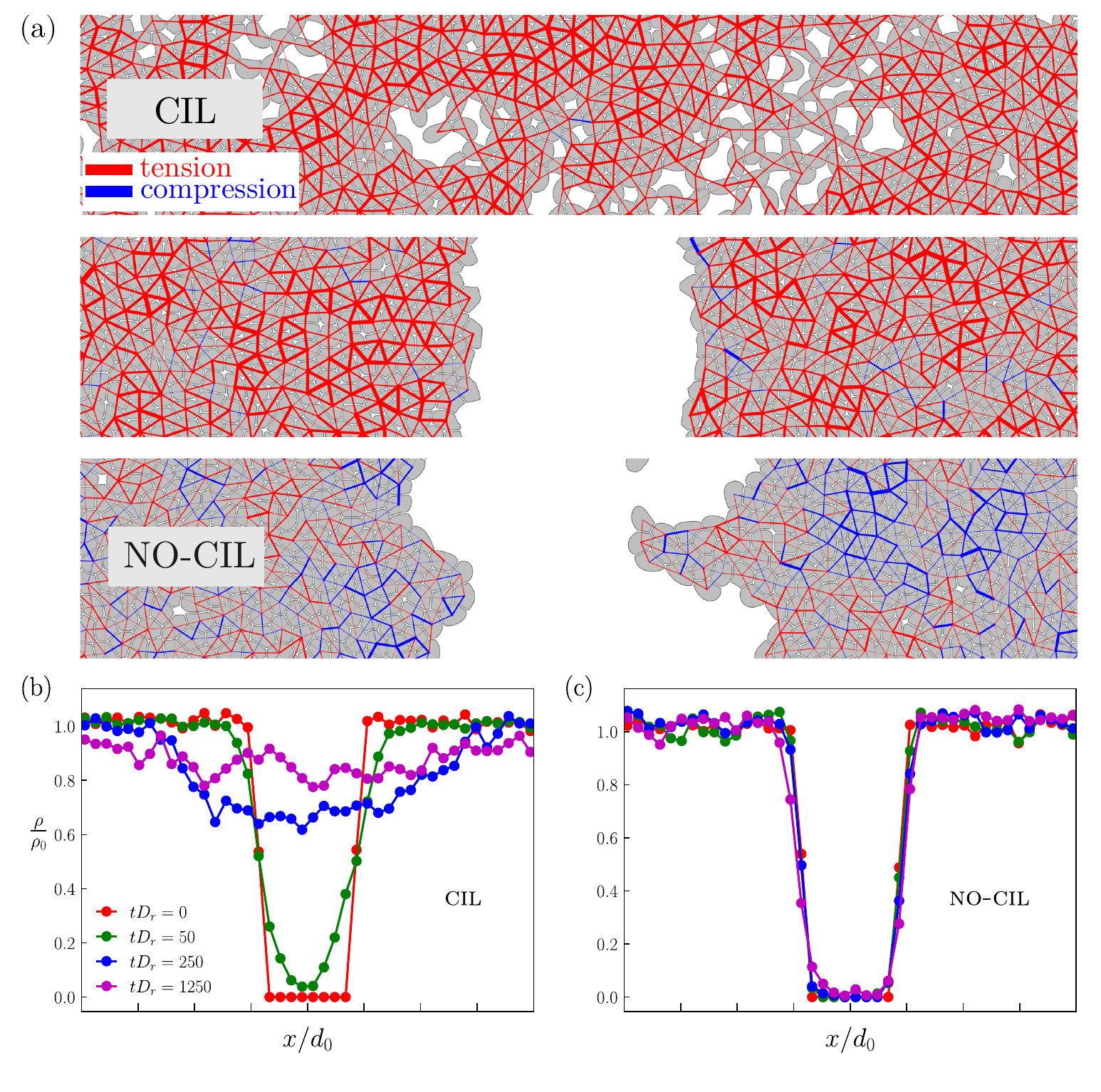}
\caption{
Evolution of a scratched array configuration at $\tilde f = 2$. Panels are as in Fig.~\ref{fig:healing_large}.
In this scenario, the system heals with \textsc{cil} while it stays phase separated in its absence.
\label{fig:healing}
}
\end{figure*}

Fig.~\ref{fig:healing} illustrates the healing dynamics a $\tilde f = 2$. 
At this $\tilde f$, the system is in the homogeneous region of the phase diagram with \textsc{cil}, in the coexistence region with \textsc{no-cil}.
Panels (a, top) and (b) clarify that with \textsc{cil} the system evolves towards a confluent steady state, even if this state occurs on a time we are unable to access numerically.
Conversely, (a bottom) and (b) clearly demonstrate that with \textsc{no-cil}, the system remains in a phase-separated configuration.

~\newpage
\section{Force chains}
\begin{figure*}[h!]
\centering
\includegraphics[width=0.9\textwidth]{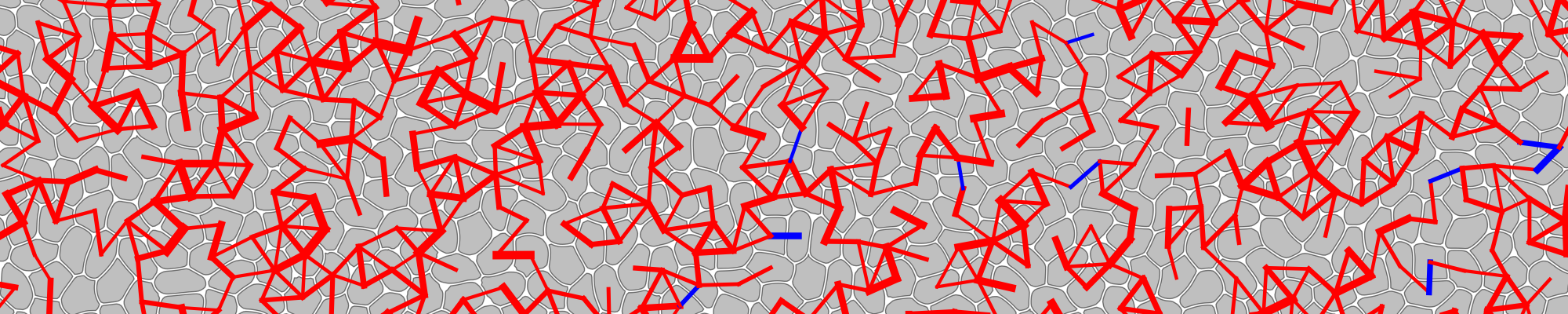}
\includegraphics[width=0.9\textwidth]{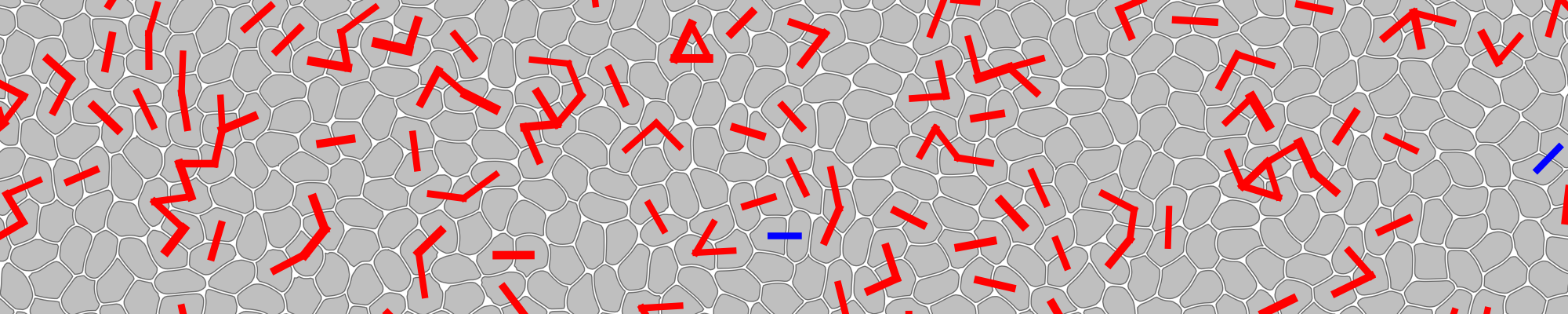}
\caption{
Final healed configuration at large $\tilde f$, as in Fig.~\ref{fig:healing_large} (top panel).
Here we illustrate exclusively the forces with a magnitude $|{\bf f}|$ larger than the average one $\langle |{\bf f}| \rangle$ (top panel), and larger than $2\langle |{\bf f}| \rangle$ (bottom panel).
\label{fig:forcechains}
}
\end{figure*}

To obtain a better insight into the geometrical properties of the force chains, we focus on the forces with a magnitude larger than some threshold.
We perform this study focussing on the final configuration of the large $\tilde f$ healed system in Fig.~\ref{fig:forcechains}.
In the top panel, we display forces with a magnitude $|{\bf f}|$ greater than the average, while in the bottom panel, we show only forces exceeding twice the average magnitude $2 \langle |{\bf f}| \rangle$.
The red colour indicates that the forces are predominantly attractive.
This figure highlights that large forces are arranged in elongated or one-dimensional structures, demonstrating their spatial heterogeneities and drawing a parallel to the force chains observed in systems of stiff particles interacting via repulsive forces. 
However, unlike in repulsive systems where force chains are primarily linear, these chains of tensile forces exhibit a high degree of bending.

